\documentclass[12pt]{article}

\usepackage[english]{babel}
\usepackage{rotating}
\usepackage{epsfig}
\usepackage{color}
\usepackage{colordvi}
\begin{document}
\date{1$^{st}$ March 2002}
\title{EIC Detector Studies
\thanks{Presented at EIC Workshop, 28 February -- 2 March, 2002, BNL}}
\author{J. Chwastowski\\
Henryk Niewodnicza\'nski Institute  of Nuclear Physics,\\ 
Cracow, Poland}
\maketitle
\begin{abstract}
The Yale '2000 Workshop detector model is presented. 
A short summary of the Interaction Region Group of the
EIC Accelerator Workshop is given.
\end{abstract}
The Yale '2000 and present workshops extensively reviewed the physics programme
for the Electron -- Ion Collider (EIC). Also at Yale and during the EIC 
Accelerator Workshop (EICAW) possible options for the machine and the 
Interaction Region (IR) were discussed.

At Yale a model of a generic detector and the IR layout were proposed and
discussed \cite{witekyale,yale}.
The design criteria were:
\begin{itemize}
\item common detector for $ep$, $eA$, $pp$ and $pA$ collisions,
\item reconstruction of ``whole'' $ep$ and $eA$ event,
\item precise luminosity monitoring and control of radiative corrections,
\item good detector quality in the fragmentation regions,
\item minimal interference with the existing IR optics,
\item use of the existing RHIC lattice in the spectrometer design,
\item possibility of a polarized electron beam.
\end{itemize}
These requirements are based on the H1 and ZEUS experiences and on the possible
range of the machine parameters \cite{erhic}.\\
The electron -- nucleus interaction leads to the production of jets, 
de-excitation gammas, ``wounded'' (struck by a probe) nucleons, re-scattered 
nucleons or heavier fragments.
\begin{figure}[hbt]
\begin{center}
\epsfig{file=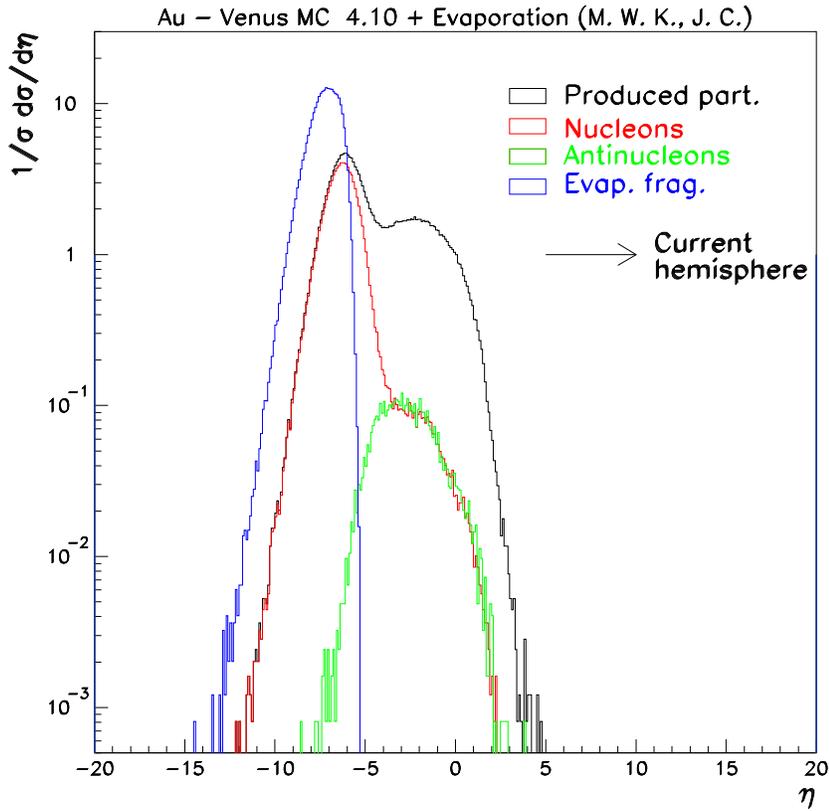,width=12cm,height=12cm}
\caption{Finale state particle pseudorapidity distribution for $e$(10 GeV) -- 
$Au$(100 GeV/A) collisions.}
\label{fig:etadist}
\end{center}
\end{figure}
Figure \ref{fig:etadist} shows the pseudorapidity ($\eta$) spectrum of the 
final state particles for $e$(10 GeV) -- $Au$(100 GeV/A) collisions. The ZEUS 
and H1 detectors at HERA prior to the machine and the detectors upgrade covered
roughly the range of (-5;5). 
Forward detectors like the Roman Pots system or the forward neutron counter
were able to cover some part of the spectrum. However, their acceptance was 
limited. 
In the electron (current) direction, the bremsstrahlung and initial state QED
radiation were measured in the photon counter. The leptons scattered at small 
angles were detected in the electron taggers.
It is obvious from Fig. \ref{fig:etadist} that to achieve a ``complete'' event 
measurement an extended ``nucleus/hadron spectrometer'' is necessary.
\begin{figure}[hbt]
\begin{center}
\epsfig{file=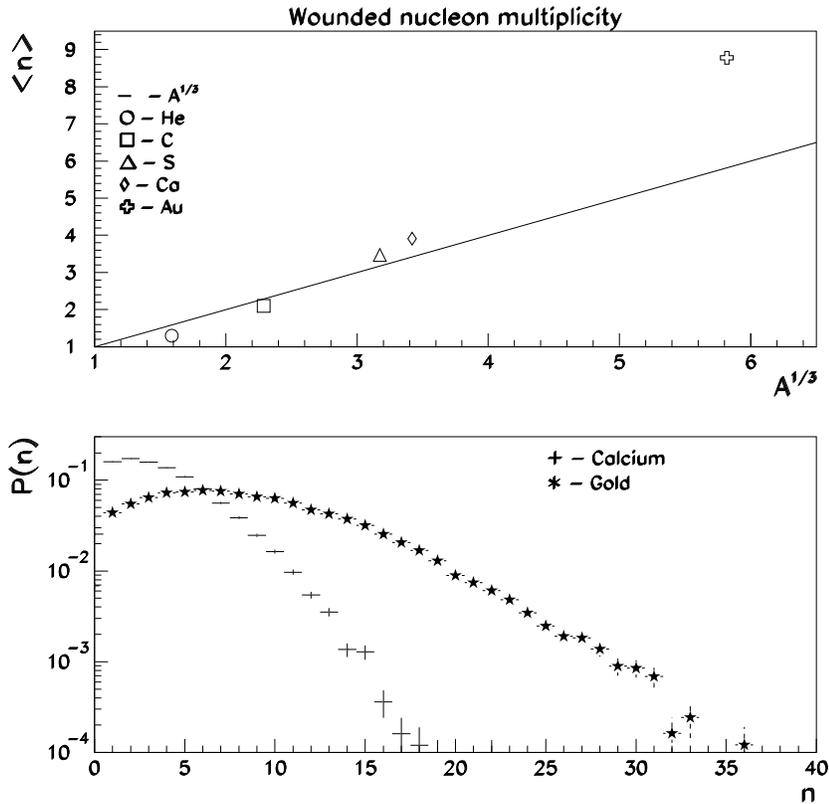,width=12cm,height=12cm}
\caption{Wounded nucleon multiplicities.}
\label{fig:nucl}
\end{center}
\end{figure}
The optimal choice of the nuclei species would cover uniformly the range of 
$R \approx A^{1/3}$. Figure \ref{fig:nucl} shows the wounded nucleon average 
multiplicity as a function of $A$ and compares the wounded 
nucleon multiplicity distributions in case of $eCa$ and $eAu$ interactions.
The average multiplicity is approximately proportional to $A^{1/3}$. 
Comparison of the distributions for calcium and gold shows that the large 
multiplicity tail in $eCa$ collisions can simulate gold-like conditions.

\begin{figure}[hbt]
\begin{center}
\epsfig{file=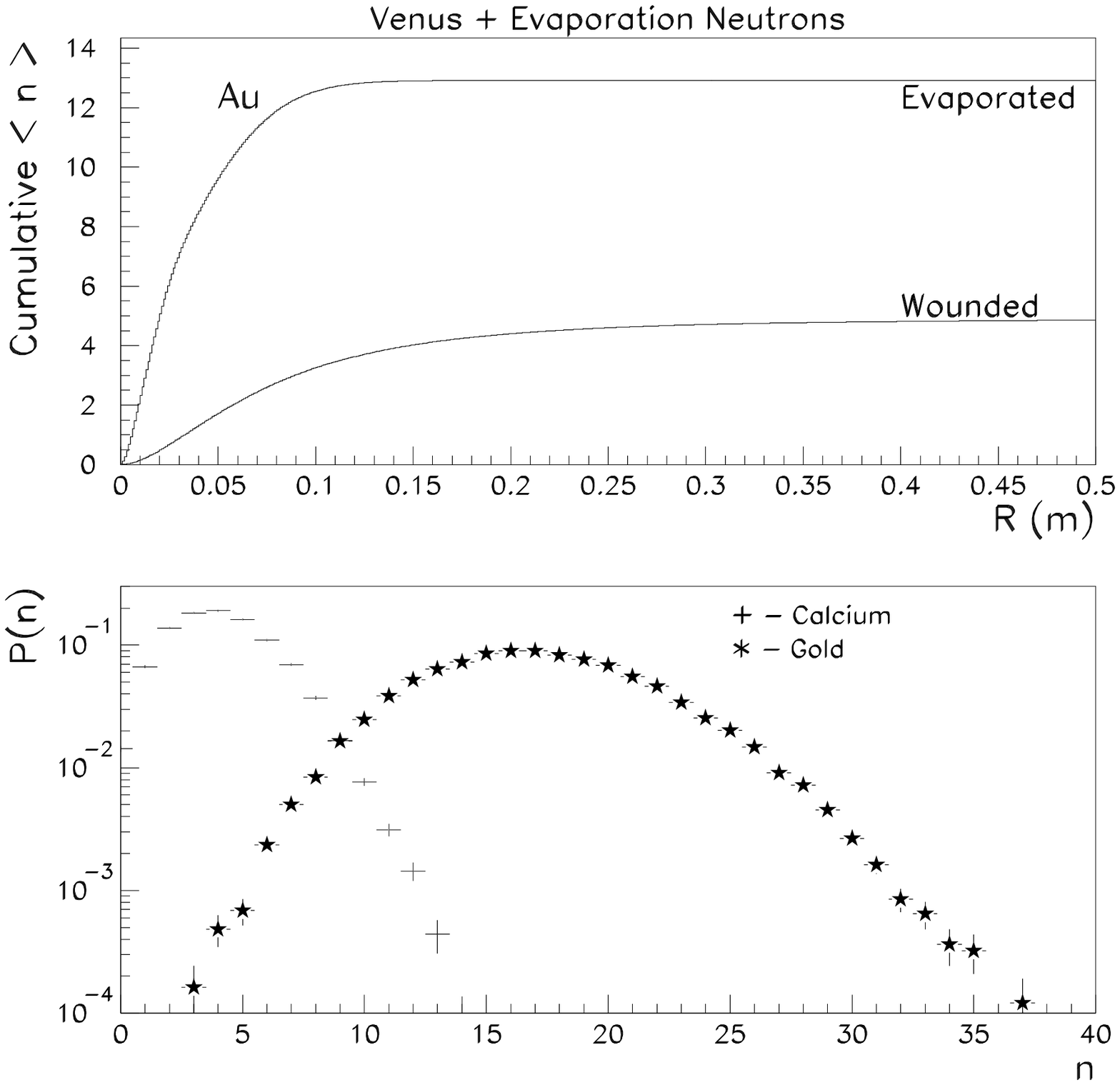,width=12cm,height=12cm}
\caption{}
\label{fig:zdc}
\end{center}
\end{figure}
\begin{figure}[hbt]
\begin{center}
\epsfig{file=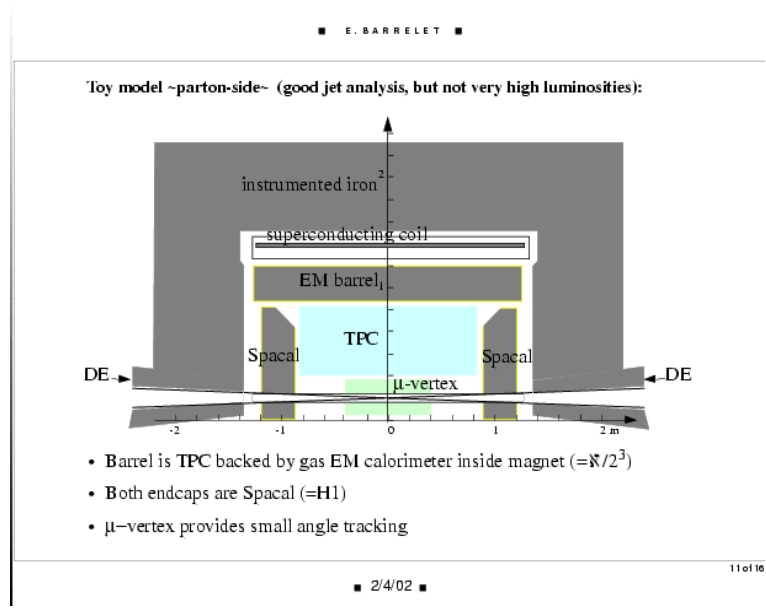,width=10cm,height=8.cm}
\caption{Yale '2000 detector and IR model. Central detector layout.}
\label{fig:cent}
\end{center}
\end{figure}
\begin{figure}[hbt]
\begin{center}
\epsfig{file=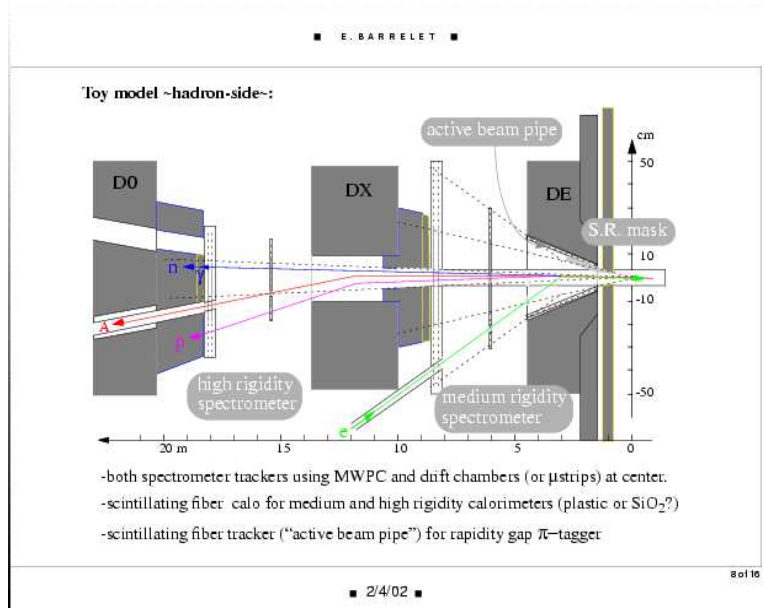,width=10cm,height=8.cm}
\caption{Yale '2000 detector and IR model. Hadron side layout.}
\label{fig:Aarm}
\end{center}
\end{figure}

In Fig. \ref{fig:zdc} the cumulative multiplicity of neutrons as a function
of the distance from the beam direction is shown. The multiplicity saturates at
the distance of about 15 -- 20 cm. This show that the enlarged Zero Degree 
Calorimeter will play important role in the inelastic event tagging.
\begin{figure}[hbt]
\begin{center}
\epsfig{file=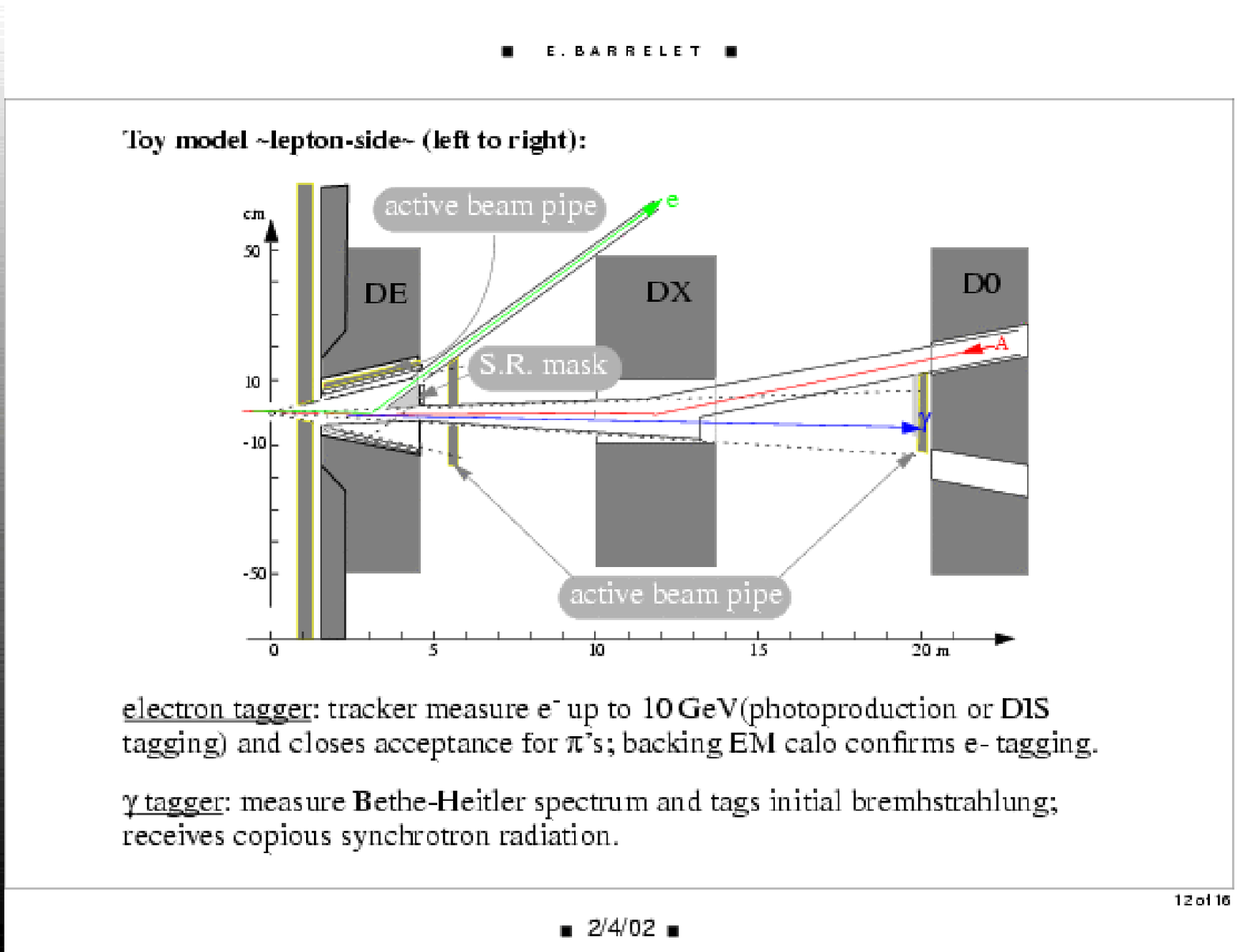,width=10cm,height=8.cm}
\caption{Yale '2000 detector and IR model. Electron side layout.}
\label{fig:earm}
\end{center}
\end{figure}
The proposed model detector (reviewed in EICAW by E. Barrelet, 
Figs.\ref{fig:earm}--\ref{fig:Aarm}) was divided 
into three parts further subdivided into the detector components:
\begin{itemize}
\item hadron side:
 \begin{itemize}
  \item Roman pots: diffractive scattering on beam,
  \item high rigidity spectrometer: EM calorimeter for nuclear 
        $\gamma$($\pi^{0}$); hadron calorimeter for measuring evaporation 
        neutrons and for p and heavier fragments identification; tracking for 
        measuring evaporation protons and heavier fragments,
  \item medium rigidity spectrometer: EM calorimeter for $\gamma$($\pi^{0}$); 
	hadron calorimeter to measure wounded neutrons, protons and for 
        identification of heavier fragments; tracking system to measure  
        $\pi^{\pm}$, protons and fragments;
  \item rapidity gap tagger: to close the acceptance for charged particles and
        to tag diffractive events;
 \end{itemize}
\item parton side is a compact central detector:
 \begin{itemize}
  \item the tracking chamber and the  EM calorimeter inside a magnet -- 
        $\aleph$/8,
  \item the Spacal (H1) type end-caps,
  \item instrumented iron,
  \item $\mu$-vertex providing small angle tracking.
 \end{itemize}
\item lepton side:
 \begin{itemize}
  \item instrumented beam pipe for electron tagging
  \item photon counter for the luminosity and radiative correction control,
        electron beam diagnostics and monitoring.
 \end{itemize}
\end{itemize}

\begin{figure}[hbt]
\begin{center}
\epsfig{file=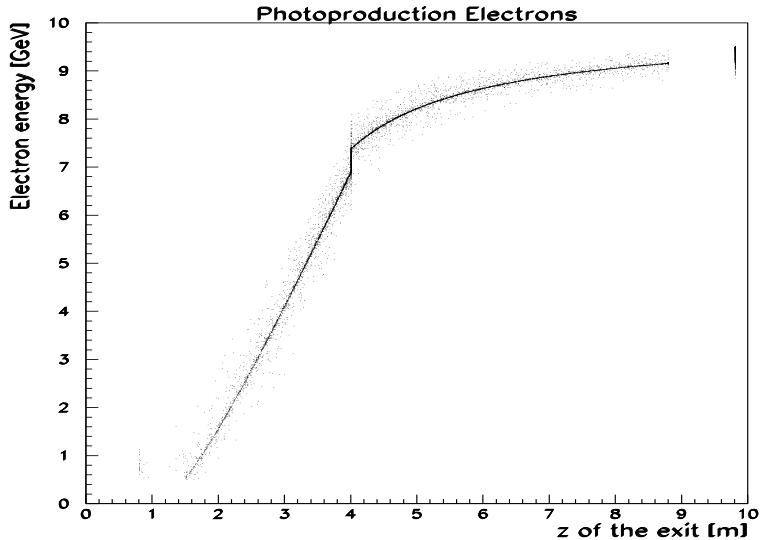,width=11cm,height=8cm}
\caption{The scattered electron energy -- position correlation.}
\label{fig:correz}
\end{center}
\end{figure}

\begin{figure}[hbt]
\begin{center}
\epsfig{file=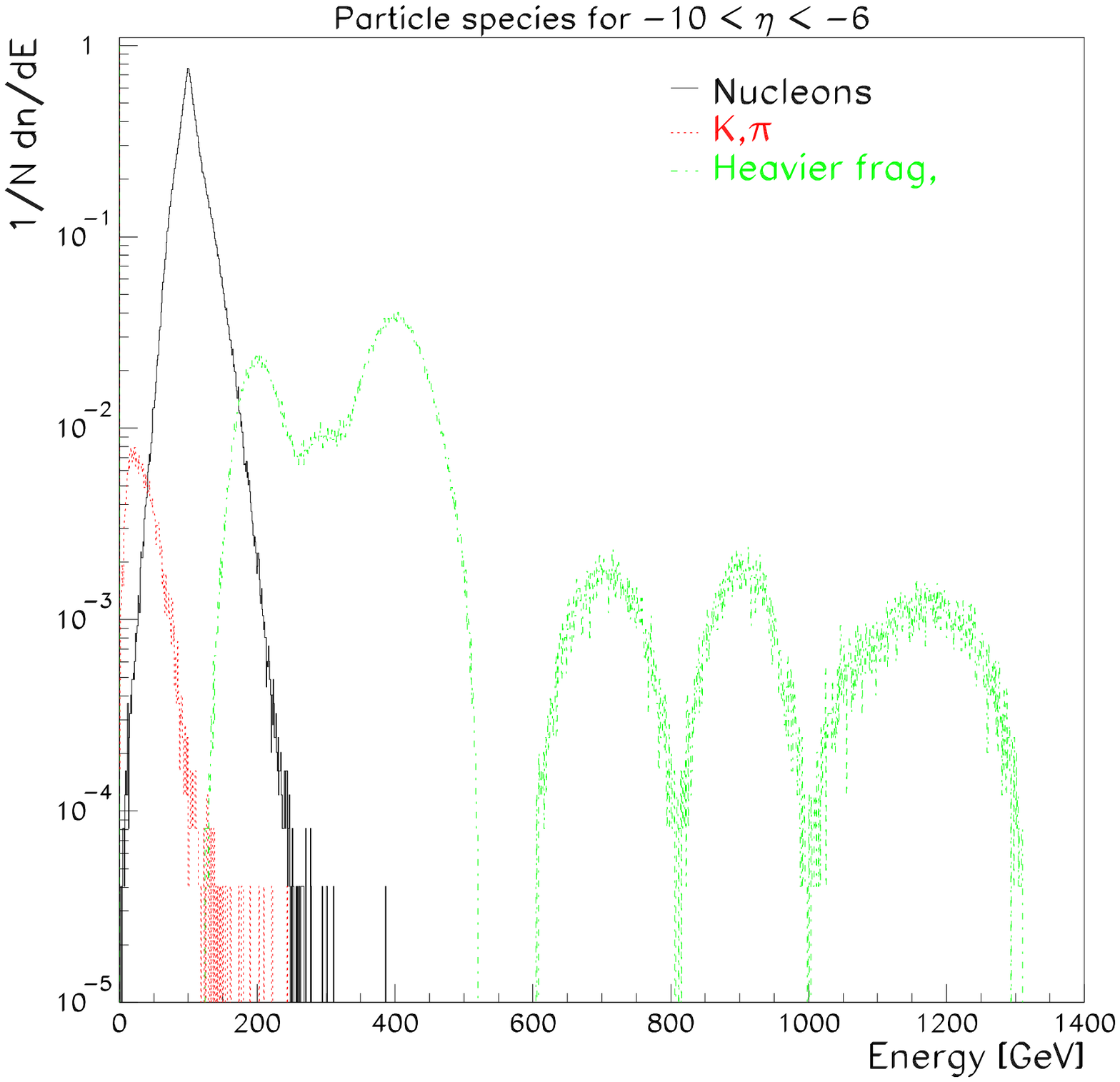,width=11cm,height=8cm}
\caption{Energy distribution of different particle species for 
$-10 < \eta < -6$.}
\label{fig:seleta1}
\end{center}
\end{figure}

As can be seen from Figs.\ref{fig:earm}--\ref{fig:Aarm} the electrons are early
separated from the ions by the $DE$ magnet. Magnet's parameters allow for the 
spin rotation and make it possible to bypass the $DX$ magnet with the electron 
beam pipe. The $DE$ magnet consists also a part of the spectrometer (medium 
rigidity) in the hadron direction. In the electron direction it is used to 
analyze the scattered electron energy. As an example a correlation between the
$z$  coordinate of the electron exit point (from the beam pipe) and its energy
is shown in Fig. \ref{fig:correz}. One can observe that the position determines
the electron energy. The beam angular divergence will broaden the correlation 
curve. The exit position can be measured with help of the active beam pipe.
For example with layers of the fibers. At the zero angle the gamma counter 
is placed. Its applications are mentioned above. In the central detector good 
energy measurement is achieved by the use of ``good resolution'' calorimeters  
combined with tracking. In addition, the $\mu$-vertex will help in the 
interaction vertex position determination and small angle tracking. The energy 
distribution of the final state particles emitted with $\eta < -6$ is shown in 
Fig. \ref{fig:seleta1}. A clear energy quantisation is seen for the nuclear 
fragments. The fragment energy determines its composition. 

The level of the synchrotron radiation with critical energy about 60 keV is a 
drawback of the Yale model.On the entry into the detector (hadron side), it may
deteriorate  the active beam pipe and the central tracking performance. On the 
exit (electron side), it will influence the lepton measurements leading to 
increased pedestals' widths. It also forces the use of the synchrotron 
radiation filters in front of the gamma counter.

\begin{figure}[hbt]
\begin{center}
\epsfig{file=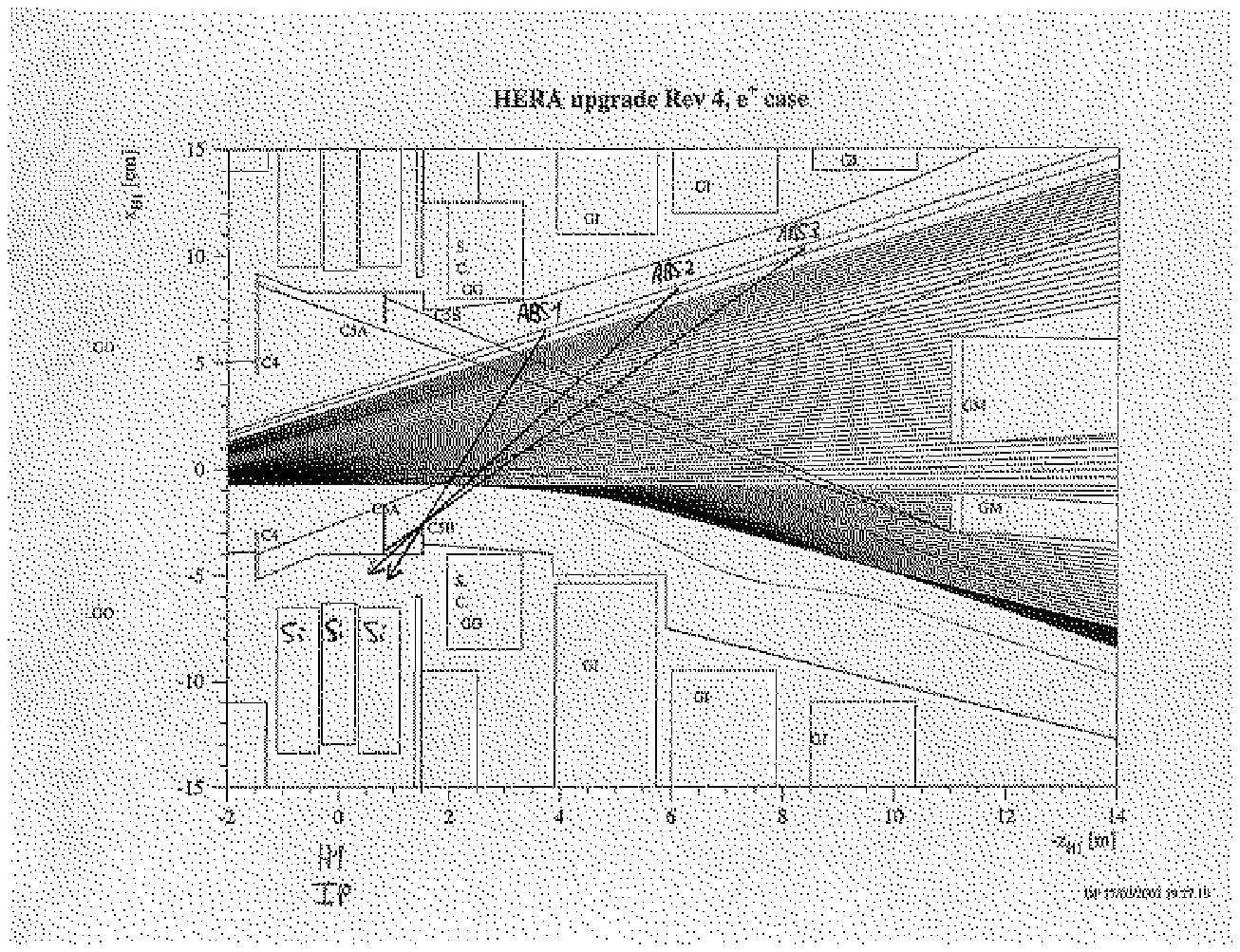,width=10cm,height=8.cm}
\caption{Synchrotron radiation in the HERA IR.}
\label{fig:daniel2}
\end{center}
\end{figure}

The HERA IR region was discussed by U. Schneekloth. To achieve high luminosity 
the final focus magnets were introduced into the central detector. This, on one
side, reduces the detector acceptance and on the other, leads to high 
synchrotron radiation levels with about $10^{18}$ photons per second. The 
calculations  shown by D. Pitzl show that a reduction factor of $10^{10}$ on
the number of photons is needed. Partially, it was achieved with help of the 
synchrotron radiation collimators (Fig. \ref{fig:daniel2}).

\begin{figure}[hbt]
\begin{center}
\epsfig{file=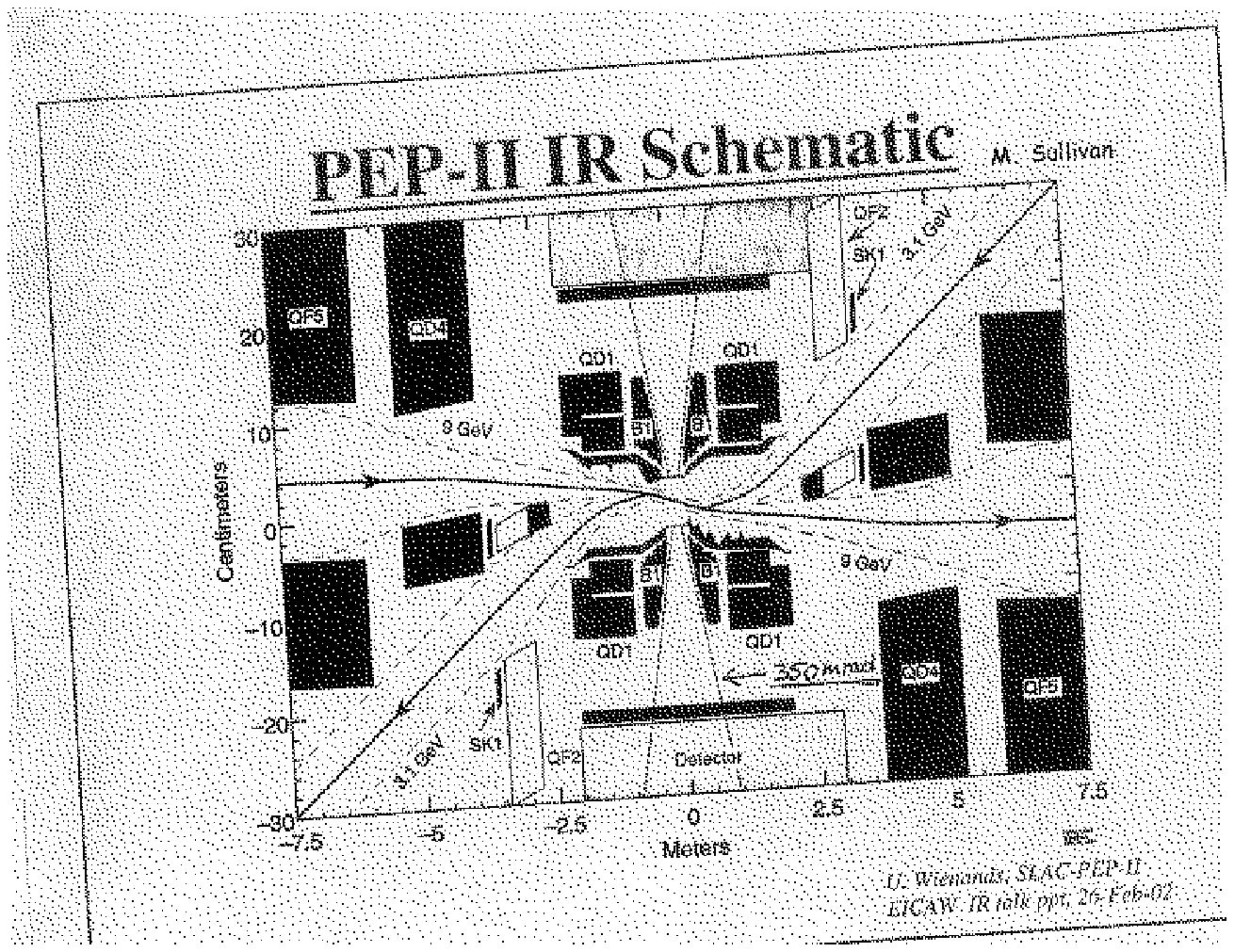,width=10cm,height=8.cm}
\caption{PEP IR.}
\label{fig:pep}
\end{center}
\end{figure}

\begin{figure}[hbt]
\begin{center}
\epsfig{file=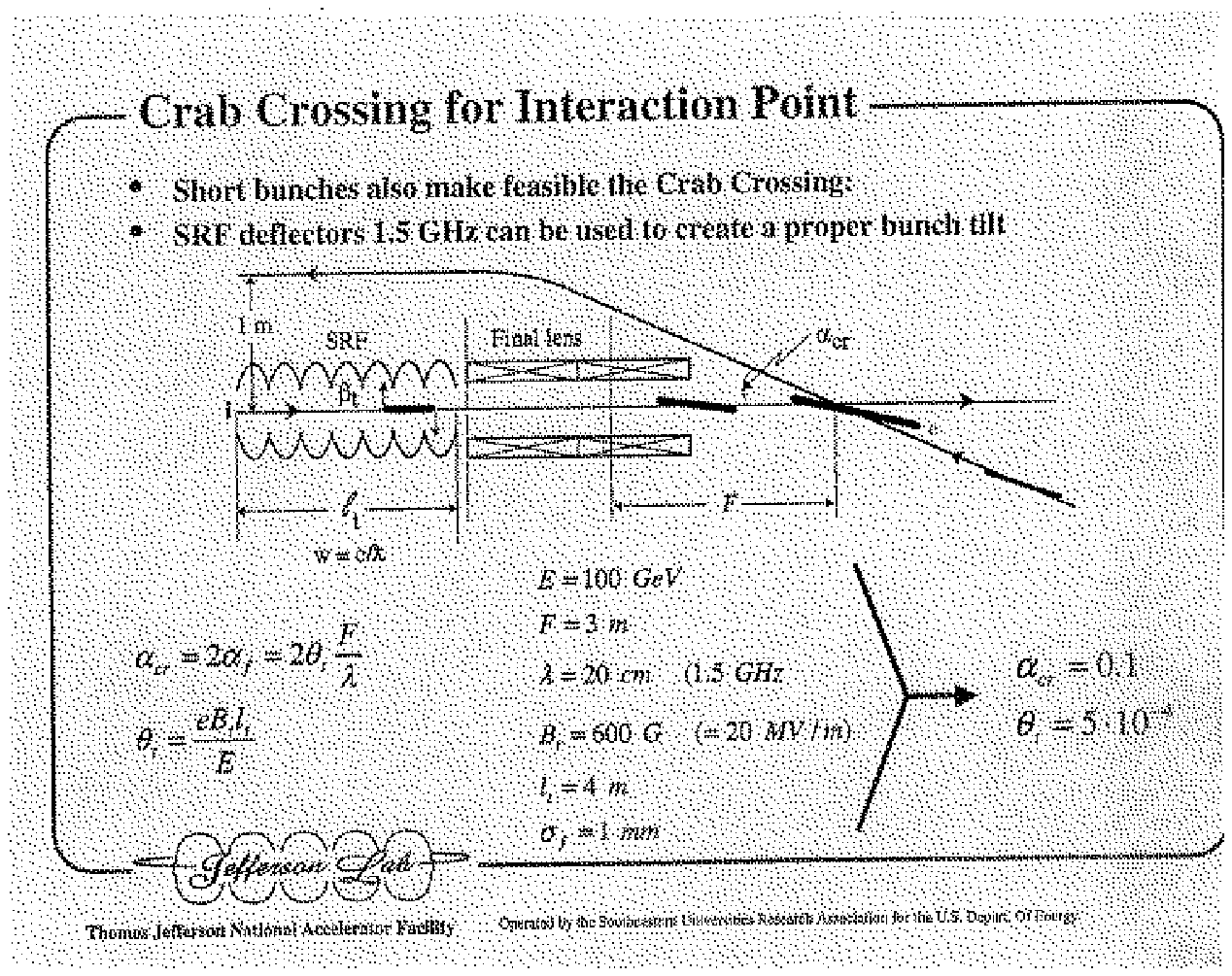,width=10cm,height=8.cm}
\caption{Crab Crossing in the IR.}
\label{fig:crab}
\end{center}
\end{figure}

An existing solution for the early separation of the beams was discussed by 
U. Wienands and is shown in Fig. \ref{fig:pep}. The price paid for the very
high luminosity in PEP are the magnets which are very close to the nominal 
vertex and the dead area up to 20$^{o}$ away from the beam direction. One 
should note that this limitation of the detector acceptance is of less 
importance in case of the$\cal B$-factory. Such a solution may play a very 
important role in case of extreme luminosities as proposed in the Crab 
Crossing option (Fig. \ref{fig:crab}) for the IR.

\begin{figure}[hbt]
\begin{center}
\epsfig{file=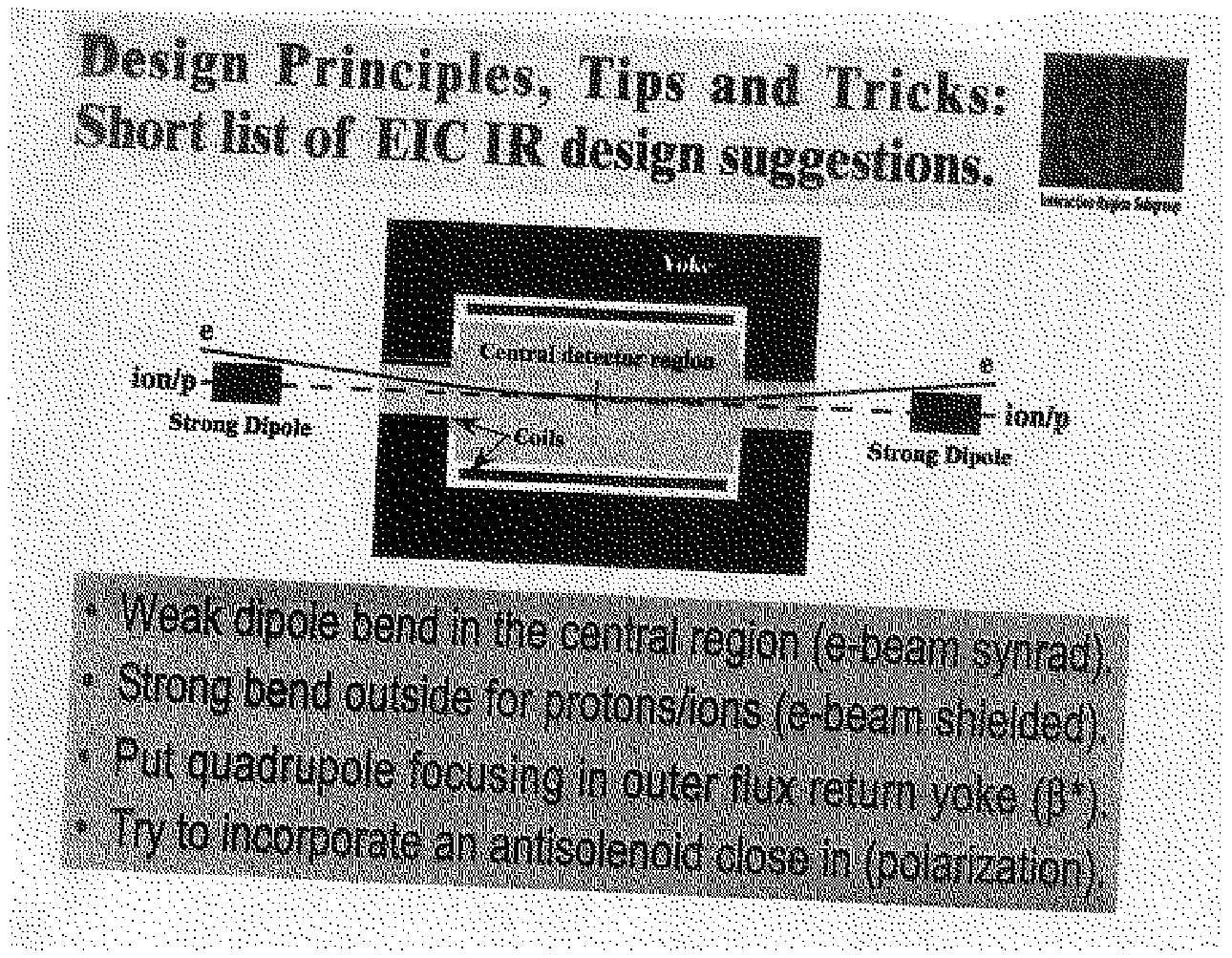,width=10cm,height=8.cm}
\caption{Early separation of beams in Parker's proposal.}
\label{fig:b1}
\end{center}
\end{figure}

\begin{figure}[hbt]
\begin{center}
\epsfig{file=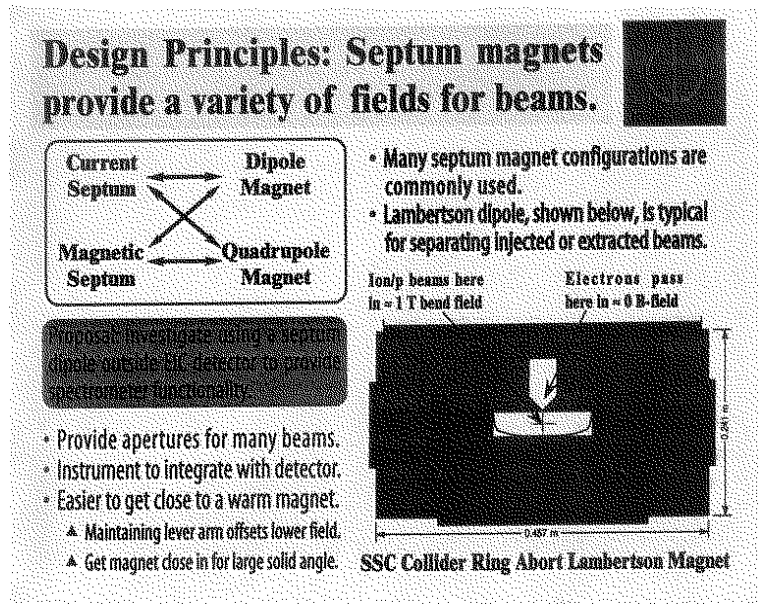,width=10cm,height=8.cm}
\caption{Lambertson magnet layout.}
\label{fig:b3}
\end{center}
\end{figure}

An improved solution for the IR was shown by B. Parker (Figs. \ref{fig:b1}--
\ref{fig:b3}). He proposes to achieve an early separation of the beams by means
 of a small dipole component added to the solenoidal central field. This 
diminishes the synchrotron radiation influence (critical energy about 6 keV). 
Further, the septum magnet (in this case the Lambertson magnet, warm type) is 
introduced next to the central detector. The electron beam passes the field 
free zone while ions traverse a 1 T dipole field. This option allows for the 
active beam pipe and provides the spectrometer functionality. One should also 
note that this option offers the head-on collisions and that all three beams 
are present in the interaction region.

Summarizing there is no doubt that a good design of the IR and that of the 
detector are very closely connected. It has to emerge from the co-operation of
the machine and the experimental physicists. It has to be checked whether the 
current designs are optimal or they should be further improved or one has to 
look for anew, different designs. One should note that the ratio of the lepton
 to the ion energies has a large impact on the detector design. So has the 
lepton energy tunability and possible luminosity value. 
Obviously one should look for the hermetic option of the detector, with the 
machine lattice used for the spectrometric measurements. The ``complete event''
option is of great value and it makes the coverage a wide range of physics 
processes feasible. The bremsstrahlung is a good candidate for a precise
luminosity measurement both on- and off-line, and for the fast on-line lepton
beam diagnostics.


\clearpage


\end{document}